\documentclass[preprint,trackchanges]{aastex63}

\pdfoutput = 1

\usepackage{changepage}

\usepackage{amsmath,amssymb,amsfonts}
\usepackage{tensor,slashed,paralist,cases,mathrsfs}

\usepackage{graphicx}

\usepackage{cancel,soul,xcolor}

\colorlet{RED}{red}
\colorlet{BLUE}{blue}
\colorlet{BROWN}{brown}

\begin{document}


\title{Diffusion of Cosmic Antiprotons Generated throughout the Dark Matter Halo \\
--- A Semi-Analytical Solution for a Linear Galactic Wind Model}

\author{Kwei-Chou Yang}
\email{kcyang@cycu.edu.tw}

\affiliation{Department of Physics and Center for High Energy Physics, Chung Yuan Christian University, \\
200 Chung Pei Road, Taoyuan 32023, Taiwan}


\begin{abstract}
The public GALPROP code gives a fully numerical solution for the spectrum of cosmic rays propagating through a linear Galactic wind directed outwards from the Galactic plane in the framework of a cylindrical diffusion model. Alternatively, for the linear Galactic wind case, we present a semi-analytical solution of the two-zone cylindrical model which describes the spectrum of cosmic-ray antiprotons produced from the primary sources throughout the dark matter halo. 
 While the secondary antiprotons can be generated by, {\it e.g.},  the GALPROP WebRun, consistently using this semi-analytical solution for the primary ones would  be  helpful to build a suited dark matter (DM) model, which may be sophisticated, and then to perform a statistical analysis when confronting with data. 
As an example, using the obtained formula, we study the possible DM signal, through the channel $\text{DM~DM} \to {\bar b}b$, and its constraint from the measurement of the AMS-02 antiproton-to-proton ratio. The indication of the DM signal is discussed. 
The advantage of the semi-analytical approach and its comparison with the GALPROP model are presented.
\end{abstract}
\keywords{Galactic cosmic rays (567), Cosmic ray sources (328), Dark matter (353)}



\section{Introduction}

The non-baryonic dark matter (DM) has been indicated by various cosmological observations and measurements \citep{Adam:2015rua,Ade:2015xua, Bergstrom:2000pn,Bertone:2004pz}, yet its nature still remains unknown.  The study of cosmic-ray nuclei, especially antimatter, has been intriguing particle physicists in order to look for underlying signals of DM annihilation (or decay). To this end, some experiments, {\it e.g.}, PAMELA and AMS, have been designed to measure these kinds of cosmic rays.  Using the GALPROP code\footnote{The code is available at ``http://galprop.stanford.edu".}  \citep{Strong:1998pw,Moskalenko:2001ya,Strong:2001fu,Vladimirov:2010aq}, several groups have reported that an excess of 10-20 GeV in the AMS-02 $\bar{p}/p$  spectrum \citep{Aguilar:2016kjl,Aguilar:2020} seems to indicate the presence of DM annihilating to ${\bar b} b$, with a DM mass $\sim$ 80~GeV \citep{Cuoco:2016eej,Cui:2016ppb,Cholis:2019ejx} which is close to the range of the DM mass hinted by the Galactic gamma-ray excess data. See Refs.~\citet{Yang:2017zor,Yang:2018fje,Yang:2019bvg,Yang:2020vxl} and references therein. 
However, the GALPROP-based data analysis is implemented in a fully numerical way. 
Including a generic DM model, or even as simple as a simplified DM model as done in Refs.~\citet{Cuoco:2016eej,Cui:2016ppb,Cholis:2019ejx}, into the GALPROP code to perform statistical  analyses is not trivial for ones who are not familiar with this program language.

The Galactic cosmic-ray propagation can be obtained by solving the transport equation \citep{Webb:1979,Jones:1990,Seo:1994}. Adopting a constant convective wind (or called Galactic wind) directed outwards from the Galactic disc, the diffusion equation for the cosmic-ray spectra generated from  sources lying in the Galactic dic or primary sources throughout the halo has been solved (semi-)analytically using the two-zone cylindrical model \citep{Webber:1992dks,Maurin:2001sj,Donato:2001ms,Maurin:2002ua}.  Nevertheless, for a linear convective wind case, which is adopted in the GALPROP code, only an approximately analytic solution of the two-zone model was given for sources located in the Galactic disc \citep{Taillet:2003yy} (see also Appendix~\ref{app:discrete-sources}), but not for the sources generated throughout the DM halo.

In this work, considering a {\it linear} wind case, we will present a semi-analytical solution for  a full two-zone cylindrical diffusion model of cosmic-ray antiprotons generated from the DM annihilation/decay in the halo.  
 This solution is important and can be complementary to the GALPROP. The advantage of the semi-analytical approach is that it can lead to a much faster computation than a fully numerical model, and provide more insights into the dependencies of relevant parameters through the solution formula.
The obtained formula is applicable to a generic case of cosmic-ray nuclei. For the two-zone diffusion model, the cosmic-ray diffusion is assumed to be isotropic and homogeneous within a cylinder of radius $R_c$ ($\sim 15-35$~kpc) and half-height $L$ ($\sim 3-6$~kpc) around the Galactic disc. On the other hand, the hydrogen and helium uniformly distribute in the disc of narrow half-thickness $h\sim 100$~pc, where the interactions of cosmic-ray nuclei with the interstellar medium (ISM) mostly occur. The solution that we obtained well approximates the cosmic-ray densities arising from DM annihilation in the halo, and is easily used to derive the constraints on a particular DM model.

As an example, we consider a possible DM annihilation scenario to interpret the antiproton excess although the astrophysical origins of this excess have  been proposed \citep{Boudaud:2019efq,Heisig:2020nse}.  We study the antiproton flux produced through the annihilation channel $\text{DM~DM} \to {\bar b}b$, and its constraint from the measurement of the AMS-02 antiproton-to-proton ratio.
 The cosmic-ray antiprotons, of which the flux is about $10^{-4}$ of cosmic-ray protons, are mainly produced as secondary particles from collisions of cosmic-ray primaries with interstellar matter. 
When fitting to the data, the antiproton spectrum contributed from DM annihilation is formulated from our semi-analytical solution, and then added to the secondary one which, together with the proton spectrum, is generated from the GALPROP WebRun \citep{Vladimirov:2010aq}, where a parameter set 
that can provide a good fit to data for the cosmic-ray proton, helium, carbon, and boron-to-carbon ratio spectra up to 200 GV is adopted from 
Ref.~\citet{Cholis:2019ejx}. Moreover, the systematic uncertainty of the production cross section for the secondary antiprotons \citep{diMauro:2014zea} and the solar modulation
effect when particles enter the heliosphere will be also taken into account in the fit. 

This paper is organized as follows. In Sec.~\ref{sec:model}, we start with a brief description of the transport equation which describes the diffusive propagation of cosmic-ray nuclei  convected by a {\it linear} wind outwards from the Galactic plane, where the nuclei could be produced from the  secondary or primary sources in the Galactic disc or throughout the halo. The relevant effects affecting the propagation of cosmic rays and their physical models are given in this section.
  In Sec.~\ref{sec:linear-solution}, we present a semi-analytical solution of the transport equation for the stationary two-zone diffusion model of cosmic-ray antiprotons produced by DM annihilation throughout the halo.
A solution for sources, primary or secondary, located in the Galactic disc is presented in Appendix~\ref{app:discrete-sources}.
  In Sec.~\ref{sec:simplified-dm}, using the semi-analytical solution for the antiproton spectrum produced from the simplified annihilation channel $\text{DM~DM} \to {\bar b} b$,  we present an analysis for AMS-02 data, and give a comparison of the present result with that obtained by the GALPROP approach.
    The $h$-dependence of the fit result is then discussed in Sec.~\ref{sec:h-dep}. 
The advantage of the semi-analytical approach is discussed in Sec.~\ref{sec:comp-adv}.
Finally, we summarize in Sec.~\ref{sec:summary}.

\section{The description of the transport equation in the Galaxy} \label{sec:model}

In our Galaxy, the cosmic-ray flux at the solar location $(r, z)=(r_\odot, 0)$ is given by
\begin{eqnarray}
\frac{d\Phi (E_K)}{d E_K} = \frac{v}{4\pi} \frac{d n (r_\odot, 0, E_K)}{d E_K} \,,
\end{eqnarray}
where $n$, $E_K$ and $v$ are the number density, kinetic energy and speed of the cosmic-ray particles, respectively. Here and in the following, $(r, z)$ are the Galactic cylindrical coordinates, with $r$ the Galactocentric radius, and $z$ the distance from the Galactic disc. 

When the cosmic rays travel to the detector at the Earth, we can separately use two transport equations to model their propagations in the Galaxy and in the Heliosphere. While for the latter its phenomenon is called solar modulation,  here we focus on the former.

The transport equation (also called the diffusion-convection equation),  used to model the distribution of specified cosmic-ray nuclei in our Galaxy, is an inhomogeneous linear integro-differential equation that can be written in the form of
\begin{eqnarray}
{\cal L} \sum_{j=b,{\rm h}} N^{(j)} = \sum_{j =b,{\rm h}} q^{(j)} \,,
\label{eq:diffusion}
\end{eqnarray}
where $N =N^{(b)}+N^{\rm (h)} \equiv dn/dE_K$ (which is the spectrum before entering the solar system) with $j=b$ or ${\rm h}$ denoting those from the background or DM halo, the terms on the right-hand side stand for the cosmic-ray sources, and
\begin{eqnarray}
{\cal L} = \frac{\partial }{\partial{t}} - \vec{\nabla}\cdot  (\overleftrightarrow{D} \vec{\nabla}  - \vec{V}_c )
   -  \frac{(\vec{\nabla}\cdot \vec{V}_c)}{3}  \frac{\partial}{\partial{E}} \Bigl( \frac{p^2}{E}  \Bigr)
 -  \frac{\partial}{\partial E} \bigg( -b_{\rm tot}(E)  + \beta^2 D_{pp}  \frac{\partial }{\partial E}  \bigg)  - \frac{1}{\tau_f}  - \frac{1}{\tau_r} -  \hat{q}^{ter} \,, \nonumber\\
 \label{eq:L}
\end{eqnarray}
with $b_{\rm tot} $ being the energy loss term, containing ionization  ($(dE/dt)_{\rm Ion}$) and Coulomb losses  ($(dE/dt)_{\rm Coul}$)  and diffusive reacceleration ($b_{\rm reac} =(1+\beta^2) D_{pp}/E$ by definition here).
Here $p$ and $E$ are the relativistic momentum and energy of the cosmic-ray nucleus, respectively.

We explain the parameters and notations as follows. 
We assume that in the local rest frame of cosmic rays the diffusive nuclei obey the Fick's law \citep{Fick:1855} and their spatial diffusion is isotropic, so that the diffusion tensor can be replaced by a scalar coefficient, $\overleftrightarrow{D} \to D_{xx}$, which is further taken as $D_{xx} = \beta D_0 (R/R_0)^\delta$  throughout the paper, where $\beta= p/E$, the rigidity of the cosmic-ray nucleus is given by $R=p/Z$, and the spectral index $\delta$ is related to the magnetohydrodynamic turbulence in the interstellar medium. $\vec{V}_c$, referred to as the velocity of the Galactic convective wind, will be considered to increase linearly with distance from the Galactic disc, i.e., $\vec{V}_c = V_0 z \hat{z}$, with $V_0$ the constant. Using the linear wind assumption, the third term of ${\cal L}$ describing the adiabatic energy losses from the convective wind can be lump into $b_{\rm tot}$ and denoted by
\begin{eqnarray}
b_{\rm adia} = -  \frac{V_0}{3} \, \frac{E_K (E_K +2m) }{E_K +m} \,,
\end{eqnarray}
where $E_K$ is the kinetic energy of the cosmic-ray nucleus.
For the brief forms of Coulomb energy losses ($(dE/dt)_{\rm Coul}$) \citep{Mannheim:1994sv} and ionization losses ($(dE/dt)_{\rm Ion}$) \citep{Mannheim:1994sv}, the reader can be referred to Ref.~\citet{Strong:1998pw}.  The momentum diffusion coefficient $D_{pp}$ is related to the spatial coefficient $D_{xx}$,  from the quasi-linear theory, as \citep{Seo:1994}
\begin{eqnarray}
D_{pp} D_{xx} =  \frac{4 p^2 v_a^2 }{3 \delta (4- \delta^2) (4- \delta) w} \,,
\end{eqnarray}
where $v_a$ is the Alfven speed, and $w$, characterizing the level of turbulence, is set to $1$ in the GALPROP \citep{Strong:1998pw}.
 While $1/\tau_r$ is the radioactive decay rate, $1/\tau_f$ describes the fragmentation rate of cosmic-ray nuclei due to annihilations with the interstellar matter; for the antiproton, it means the annihilation rate of $\bar{p}$ on the interstellar protons in the Galactic disc, and is approximately given by 
 \begin{eqnarray}
 1/\tau_f \simeq \beta c (n_H + 4^{2/3} n_{He} ) \sigma_{p\bar{p}}^{\rm ann} \,,
 \label{eq:antip-ann}
 \end{eqnarray}
where $n_H$ and $n_{He}$ are hydrogen and helium densities, respectively, and \citep{Tan:1983de,Groom:2000in,Moskalenko:2001ya}
\begin{eqnarray}
\sigma_{p \bar{p}}^{\rm ann} = 
\left\{ \begin{array}{lr}  661 (1 +0.0115\, E_K^{-0.774} - 0.948\, E_K^{0.0151})~{\rm mbarn},   \ \ 
        & \text{for $E_K \leq10$~GeV} \\ 
      70.86 \, [2 \, m_p (E_K+m_p)] ^{-0.56}~  {\rm mbarn} \, , &  \text{for $E_K > 10$~GeV} \end{array} \right. \; .
\end{eqnarray}
In Eq.~(\ref{eq:antip-ann}),  ``$4^{2/3}$" a geometrically-inspired factor accounting for the cross section \citep{Maurin:2002ua}.
 For antiprotons, one has to consider the last term in Eq.~(\ref{eq:L}). This term, called the tertiary component, describes that the antiprotons survive during the inelastic scatterings with the protons which are then excited to resonant states.
The tertiary contribution{\footnote{We call it to be the tertiary contribution throughout this paper as in Ref.~\citet{Cirelli:2010xx}, although this term is secondary for the case that antiprotons are generated from DM annihilation.} is given by  \citep{Donato:2001ms}
\begin{eqnarray}
\hat{q}^{ter} N^{(j)} 
     &\equiv& q^{(j) ter} \nonumber\\
     &=& \int_{E_K}^\infty  \frac{\sigma_{p\bar{p}}^{\text{non-ann}} (E_K^{\prime}) }{E_K^\prime} (n_H + 4^{2/3} n_{He} ) v^{\prime} N^{(j)} (E_K^\prime) dE_K^{\prime} 
- \sigma_{p\bar{p}}^{\text{non-ann}} (E_K) (n_H + 4^{2/3} n_{He} )  v N^{(j)}  \,,
\end{eqnarray}
where the $p$-$\bar{p}$ non-annihilation cross section is equal to its inelastic scattering cross section minus the annihilation cross section,
\begin{eqnarray}
\sigma_{p\bar{p}}^{\text{non-ann}} =  \sigma_{p\bar{p}}^{\rm ine} - \sigma_{p\bar{p}}^{\rm ann} \,,
\end{eqnarray}  
where  the  inelastic scattering cross section reads \citep{Tan:1983de,Moskalenko:2001ya}
\begin{eqnarray}
\sigma_{p \bar{p}}^{\rm ine}  = 
\left\{ \begin{array}{lr}   24.7 (1 +0.584\, E_K^{-0.115} +0.856\, E_K^{-0.566})~{\rm mbarn}\,,   & \text{for $E_K \leq 14$~GeV} \\ 
      32.2 [1+ 0.0273 \,  U+ 0.01 \, U^2  \, \theta(U) ~  {\rm mbarn}  + \sigma_{p \bar{p}}^{\rm ann} \,, ~~~ & \text{for $E_K > 14$~GeV} 
\end{array} \right. \;,
\end{eqnarray}
with $U= \ln [(m_p +E_K)/200]$.
The interactions of cosmic-ray nuclei with the interstellar medium mainly occur within the narrow Galactic disc with half-height,  $h$, of order $\sim 100$~pc $\ll L$.   In the analysis of the two-zone cylindrical model, we approximated the density of the interstellar medium to be $n(r,z) = 2 h \delta(z) n^{\rm ISM}$.
Thus, in Eq.~(\ref{eq:antip-ann}),  the fragmentation rate of antiprotons due to $p$-$\bar{p}$ annihilation can be represented as
 \begin{eqnarray}
 1/\tau_f
  \simeq 2 h\, \delta(z)\,  \Gamma_{\rm ann}
  = 2 h\, \delta(z)\,  \beta c \big (n_H^{\rm ISM} + 4^{2/3} n_{He}^{\rm ISM} \big) \sigma_{p\bar{p}}^{\rm ann} \,,
 \label{eq:antip-ann-form2}
 \end{eqnarray}
 where $n_H^{\rm ISM} \approx 0.9~\text{cm}^{-3}$ and $n_{He}^{\rm ISM} \approx  0.11~n_H^{\rm ISM} $ will be taken.

\section{Solution of the diffusion-convection equation with a linear wind velocity}\label{sec:linear-solution}

An important property of Eq.~(\ref{eq:diffusion}) is that this equation obeys the principle of superposition: ${\cal L} N^{(b)} =q^{(b)}$ and ${\cal L} N^{\rm (h)} =q^{\rm (h)}$, so that the resulting total number density spectrum can be written as $N = N^{(b)} + N^{\rm (h)}$. In other words, we can separately calculate the contributions arising from the background and DM halo.
For the case of cosmic-ray antiprotons, the background stands for those produced as secondary particles from collisions of cosmic-ray primaries with interstellar matter.

In the following calculation, we assume that the cosmic-ray nuclei, before entering the solar system, are in the steady state, i.e., $\partial N^{(j)}/ \partial t =0$, and propagate within a cylindrical region centered at the Galaxy center. The cylinder has a radius $R_c \sim 25$~kpc and half-height $L$ which, of order several kpc, is still to be determined. We adopt the so-called two-zone diffusion model in which the boundary conditions are $N^{(j)}(r=R_c, z) = N^{(j)}(r, z=\pm L)=0$, and the gas, mostly made of hydrogen and helium, uniformly concentrates within a narrow half-height $h \sim 100$~pc. Thus, the steady diffusion equation with a linear velocity of the convective wind,  directed outwards from the Galactic plane, can be approximated as 
\begin{eqnarray}
D_{xx} \left[ \frac{\partial^2}{\partial z^2} + \frac{1}{r}  \frac{\partial}{\partial r} \bigg( r \frac{\partial}{\partial r} \bigg)\right] N^{(j)}
  & -&  V_0  \frac{d}{dz} (z\, N^{(j)}) - (\Gamma_{\rm r} + 2 h \Gamma_{\rm ann} \delta(z) ) N^{(j)}  \nonumber \\
  &=& -q^{(j)}  - q^{(j) ter} - \frac{\partial}{\partial E} \bigg( -b_{\rm tot}(E) N^{(j)} + \beta^2 D_{pp}  \frac{\partial N^{(j)}}{\partial E}  \bigg) 
   \,,
   \label{eq:diffusion2form}
\end{eqnarray}
where $\Gamma_{\rm r} = 1/\tau_{\rm r}$.
 In the following, we will simply focus on the study of cosmic-ray antiprotons; thus we have $\Gamma_{\rm r} =0$, although it is still kept in the formula. Meanwhile, for physical quantities, the kinetic energy, $E_K=E-m_p$, will be used as a variable, instead of the total energy, $E$. The following derivation is also applicable to a generic case of cosmic-ray nuclei.

We represent the number density $N^{(j)}$ and source $q^{(j)}$, both of which are isotropic and satisfy the boundary conditions, $N^{(j)} (R_c,z,E_K)=0$ and $q^{(j)} (R_c,z,E_K)=0$, in terms of a complete set of orthogonal Bessel functions:
\begin{eqnarray}
N^{(j)} (r, z, E_K) &=& \sum_{i=1}^{\infty} N_i^{(j)}  (z, E_K) J_0 (\zeta_i \rho) \,,  \\
q^{(j)}(r, z, E_K) &=& \sum_{i=1}^{\infty} q^{(j)}_i (z, E_K) J_0 (\zeta_i \rho) \,,
\end{eqnarray}
where $\rho=r/R_c$ and $\zeta_i$ is the $i^{\rm th}$ zero of the Bessel function $J_0$.  While for the background (i.e., the secondary antiprotons) related to the sources in the Galactic disc, the solution can be obtained in a similar way and will be collected in Appendix~\ref{app:discrete-sources} for reference, here we focus on the calculation for the cosmic-ray density generated from the sources $q^{\rm (h)}$  in the DM halo, and then rewrite Eq.~(\ref{eq:diffusion2form}) as
\begin{eqnarray}
D_{xx} \frac{\partial^2}{\partial z^2}  N^{\rm (h)}_i
  & -&  V_0  \frac{d}{dz} (z\, N^{\rm (h)}_i) - \Bigg( \frac{D_{xx} \zeta_i^2}{R_c^2}  + \Gamma_{\rm r} + 2 h \Gamma_{\rm ann} \delta(z) \Bigg) N^{\rm (h)}_i  \nonumber \\
  &=& -q^{\rm (h)}_i  - \hat{q}^{ter} N^{\rm (h)}_i  - \frac{\partial}{\partial E_K} \bigg( -b_{\rm tot}(E_K) N^{\rm (h)}_i + \beta^2 D_{pp}  \frac{\partial N^{\rm (h)}_i }{\partial E_K}  \bigg) 
   \,.
   \label{eq:diffusion-z-dep}
\end{eqnarray}

To solve the cosmic-ray density from Eq.~(\ref{eq:diffusion-z-dep}), we first consider the  approximation  for which the energy derivatives and tertiary contribution can be neglected. Under this approximation, introducing the new variable $y$ to substitute $z$,
\begin{eqnarray}
y =\bigg( \frac{V_0}{2 D_{xx}} \bigg)^{1/2} z \,,
\end{eqnarray}
the diffusion-convection equation containing only the DM halo primary source satisfies  (for $z\lessgtr 0$) 
\begin{eqnarray}
\frac{d N_i^{\rm (h)}}{d y^2} - 2 y \frac{d N_i^{\rm (h)}}{d y} -c_i N_i^{\rm (h)} = - \frac{2}{V_0} q_i^{\rm (h)}  \,,
\label{eq:width-ann-simplified}
\end{eqnarray}
where 
\begin{eqnarray}
c_i = \frac{2 D_{xx}}{V_0} \Bigg( \frac{\zeta_i^2}{R_c^2} + \frac{\Gamma_{\rm r}}{D_{xx}} \Bigg)+2 \,.
\end{eqnarray}
The solution of this equation can be written in the form
\begin{eqnarray}
N_i^{\rm (h)}(z, E_K) = -y_p^i (y, E_K) + \alpha_i (E_K) y_1^i (y)+ \beta_i (E_K)  y_2^i (y), \label{eq:sol-1}
\end{eqnarray}
where the complementary function, composed of $y_1^i$ and $y_2^i$, has two free parameters $\alpha_i$ and $\beta_i$, which are functions of $E_K$ and can be determined by the boundary conditions, and the particular function is denoted as $y_p^i$. These functions are given by
\begin{eqnarray}
y_1^i(y) &=& y\,  {}_1F_1\left( \frac{2+c_i}{4}, \frac{3}{2}; y^2 \right) \,, \\
y_2^i(y) &=& {}_1F_1\left( \frac{c_i}{4}, \frac{1}{2}; y^2 \right) \,, \\
y_p^i (y, E_K) &=& \frac{2}{V_0} \int^y_0 G^i (y, y' ) \,  q_i^{\rm (h)} (z', E_K) \, dy' \,,
\end{eqnarray}
where  $y_1^i$ and $y_2^i$ are adopted to satisfy $y_1^i(0)=0, y_2^i(0)=1, dy_1^i(0) /dy =1$, and $ dy_2^i(0) /dy =0$, ${}_1F_1$ is the confluent hypergeometric function of the first kind, the component of the halo distribution is
\begin{eqnarray}
q_i^{\rm (h)} (z, E_K)
&=& \frac{2}{J_1^2(\zeta_i) R_c^2} \int_0^{R_c} dr \, r \, J_0 \bigg( \zeta_i \frac{r}{R_c} \bigg)  q^{\rm (h)}(r, z, E_K)  \nonumber \\
&=& \frac{2}{J_1^2(\zeta_i) R_c^2} \int_0^{R_c} dr \, r \, J_0 \bigg( \zeta_i \frac{r}{R_c} \bigg)  \Bigg( \frac{\rho_{\rm DM} (r, z')}{ \rho_\odot} \Bigg)^2
\frac{1}{2} \bigg( \frac{\rho_\odot}{m_{\rm DM}} \bigg)^2  \sum_k \langle \sigma v \rangle_k   \bigg(\frac{ d {\cal N}_{\bar p}}{d E_K} \bigg)_k  
\nonumber\\
& \equiv& \sum_k q_{i,k}^{\rm (h)} (z, E_K) \,,
\label{eq:qh-i}
\end{eqnarray}
and the Green's function is
\begin{eqnarray}
G^i(y, y') = \frac{y_1^i (y') y_2^i(y) - y_1^i(y) y_2^i (y')}{ y_1^i (y') y_2^{i\prime}(y') - y_1^{i\prime}(y') y_2^i (y') } \,.
\end{eqnarray}
Here $\langle \sigma v\rangle_k$ is the cross section of the DM annihilation into a state denoted by ``$k$", 
and  $ (d {\cal N}_{\bar p} /d E_K)_k $ is the resulting antiproton spectrum produced per DM annihilation into this state $k$ in the center-of-mass frame of DM. 

To satisfy the boundary condition $N_i^{\rm (h)}(z=L) =0$, we require, from Eq.~(\ref{eq:sol-1}), that
\begin{eqnarray}
\alpha_i = \frac{y_p^i (y_L, E_K) - \beta_i y_2^i (y_L)}{y_1^i (y_L)} \,,
\label{eq:alphai}
\end{eqnarray}
with
\begin{eqnarray}
y_L  \equiv \bigg( \frac{V_0}{2 D_{xx}} \bigg)^{1/2} L \,.
\end{eqnarray}
Second, we further take into account the corrections, resulting from energy derivatives and tertiary contribution, to the above approximation of $N^{\rm (h)}_i$ at the solar system located at $z=0$ that we concern. 
The density spectrum of the cosmic antiprotons, generated from the DM annihilation, at the solar system can be obtained to be a function of $\beta_i$:
\begin{eqnarray}
N^{\rm (h)} (r_\odot, z=0, E_K) 
&=& \sum_{i=1}^{\infty} N_i^{\rm (h)}  (z=0, E_K) J_0 \Big( \zeta_i \frac{r_\odot}{R_c} \Big)\nonumber \\
&=& \sum_{i=1}^{\infty} \beta_i J_0  \Big( \zeta_i \frac{r_\odot}{R_c} \Big)  
= \sum_k \sum_{i=1}^{\infty}  \beta_{i,k}  J_0 \Big( \zeta_i \frac{r_\odot}{R_c} \Big) \,,
\label{eq:multi-expansion}
\end{eqnarray}
where in the last line $\beta_i= \sum_k \beta_{i,k}$, of which each term results from the DM annihilation (or decay) channel, ${\rm DM~DM} \to k \to {\bar p}+ X $ (or  ${\rm DM} \to k \to {\bar p} + X $). For simplicity, in the following we will focus on the DM annihilation case. With/Without the corrections, the value of $\beta_{i,k}$ can be directly determined in the following way.

For Eq.~(\ref{eq:diffusion-z-dep}), substituting $N_i^{(h)}$ by its expansion form given in Eq.~({\ref{eq:multi-expansion}), and then performing the integration of this equation\footnote{
The key point for the calculation is that, after integration with respect to $z$ from $-h$ to $+h$, the first term of the left-hand side of Eq.~(\ref{eq:diffusion-z-dep})  is given by 
\begin{eqnarray}
2D_{xx} \frac{d}{dz} N^{\rm (h)}_i (z) |_{z=h} = -\int_{-h}^{h} q^{\rm (h)}_i (z') dz' + 2D_{xx} \bigg( \frac{V_0}{2D_{xx} }\bigg)^{1/2} \alpha_i 
+ 2h \beta_i \Bigg( V_0 +  \frac{D_{xx} \zeta_i^2}{R_c^2}  + \Gamma_{\rm r}  \Bigg)
+{\cal O}(h^2) \,.
\end{eqnarray}
} with respect to $z$ from $-h$ to $+h$, where the narrow range covers the interstellar gas,  we obtain
\begin{eqnarray}
\beta_{i,k} \left[ 2 h \Gamma_{\rm ann} +2 D_{xx} \bigg( \frac{V_0}{2D_{xx}} \bigg)^{1/2}  \frac{y_2^i(y_L)}{y_1^i(y_L)} \right]
 =& 2& D_{xx} \bigg( \frac{V_0}{2D_{xx}} \bigg)^{1/2}  \frac{y_p^{i,k} (y_L)}{y_1^i(y_L)} \nonumber\\
   & - & 2h \frac{\partial}{\partial E_K} \left( b_{tot} \beta_{i,k} - \beta^2 D_{pp} \frac{\partial \beta_{i,k} }{\partial E_K} \right) 
   + 2h q_{i,k}^{\rm ter} (r_\odot, 0, E_K) \,, \nonumber\\
  \label{eq:betaik}
\end{eqnarray}
where 
\begin{eqnarray}
y_p^{i,k} (y) &=& \frac{2}{V_0} \int^y_0 G^i (y, y' ) \,  q_{i,k}^{\rm (h)} (z', E_K) \, dy' \,, \\
q^{\rm ter}_{i,k}
     &=&\int_{E}^\infty  \frac{\sigma_{p\bar{p}}^{\text{non-ann}} (E_K^{\prime}) }{E_K^\prime} (n_H + 4^{2/3} n_{He} ) v^{\prime} \beta_{i,k} (E_K^\prime) dE_K^{\prime} 
- \sigma_{p\bar{p}}^{\text{non-ann}} (E_K) (n_H + 4^{2/3} n_{He} ) \, v  \, \beta_{i,k}  \,.
\end{eqnarray}
Numerically, we can apply iterative procedure to Eq.~(\ref{eq:betaik}) to get the solution of $\beta_{i,k}$. We rename $\beta_{i,k}$ on the right-hand side and left-hand side of Eq.~(\ref{eq:betaik}) to be $\beta_{i,k}^{(n)}$ and $\beta_{i,k}^{(n+1)}$, respectively. $\beta_{i,k}^{(0)}$, which corresponds to the zeroth approximation when neglecting  the energy loss term, reacceleration and tertiary contribution, can be obtained from Eq.~(\ref{eq:betaik}), given by
 \begin{eqnarray}
\beta_{i,k}^{(0)} 
 = \frac{\bar{y}_p^{i} (y_L) }{A_i \, y_1^i (y_L)} 2 D_{xx} \bigg( \frac{V_0}{2D_{xx}} \bigg)^{1/2} 
 \frac{1}{2} \bigg( \frac{\rho_\odot}{m_{\rm DM}} \bigg)^2   \langle \sigma v \rangle_k  \,  \bigg(\frac{ d {\cal N}_{\bar p}}{d E_K} \bigg)_k  \,,
 \label{eq:beta0-ik}
\end{eqnarray}
with
\begin{eqnarray}
\bar{y}_p^{i} (y_L)  
             & =&\frac{4}{J_1^2(\zeta_i) } \frac{1}{V_0 R_c^2} \int_0^{R_c} dr \, r \, J_0 \bigg( \zeta_i \frac{r}{R_c} \bigg)
\int_0^{y_L} G^i (y_L, y' ) dy'  \Bigg( \frac{\rho_{\rm DM} (r, z')}{ \rho_\odot} \Bigg)^2 \,, \\
A_i^{-1} &=& \frac{y_1^i (y_L)}{2D_{xx}} \frac{1}{  \big( \frac{V_0}{2 D_{xx}} \big)^{1/2}  y_2^i(y_L) + \frac{h \Gamma_{\rm ann}}{ D_{xx}} y_1^i(y_L)} \,.
\label{eq:Ainv}
\end{eqnarray}
 In this paper,  the spectra $ (d {\cal N}_{\bar p} /d E_K)_k $ are generated from the PPPC4DMID code\footnote{This code is  available at the website: ``http://www.marcocirelli.net/PPPC4DMID.html".}  \citep{Cirelli:2010xx,Ciafaloni:2010ti}, which, containing the electroweak corrections, was calculated using PYTHIA 8.135 \citep{Sjostrand:2007gs}.  
Using the Mathematica, which is a symbolic computation program, one can easily obtain the iterative solution.

The antiproton flux generated from the DM annihilation at the location of the Sun, but before entering the solar system,  thus reads
\footnote{
For the DM decay, the result can be obtained by performing the following replacement in Eqs.~(\ref{eq:qh-i}), (\ref{eq:beta0-ik}), and (\ref{eq:DM-antiproton}), 
\begin{eqnarray}
\frac{1}{2} \Big( \frac{\rho_\odot}{m_{\rm DM}} \Big)^2   \langle \sigma v \rangle_k \,   \Big(\frac{ d {\cal N}_{\bar p}}{d E_K} \Big)_k 
\to  \frac{\rho_\odot}{m_{\rm DM}} \Gamma_k \,   \Big(\frac{ d {\cal N}_{\bar p}}{d E_K} \Big)_k^{\rm decay} 
 \,,
\label{eq:DM-decay}
\end{eqnarray}
where on the right-hand side, $\Gamma_k$ is the DM decay rate into the final state $k$, and $(d {\cal N}_{\bar p} / d E_K)_k^{\rm decay}$ is the antiproton spectrum produced per DM decay to the $k$ state.
}
\begin{eqnarray}
\frac{d\Phi_{\bar p}^{\rm DM} (E_K)}{d E_K} = \frac{v}{4\pi} \frac{d n_{\bar p}^{\rm DM} (r_\odot, 0, E_K)}{d E_K} \,,
\label{eq:flux}
\end{eqnarray}
where
\begin{eqnarray}
\frac{d n_{\bar p}^{\rm DM} (r_\odot, 0, E_K)}{d E_K} \,
\big(\equiv N^{\rm (h)} (r_\odot, z=0, E_K) \big)
= \frac{1}{2} \bigg( \frac{\rho_\odot}{m_{\rm DM}} \bigg)^2  \sum_k \langle \sigma v \rangle_k \,   \bigg(\frac{ d {\cal N}_{\bar p}}{d E_K} \bigg)_k   
R_k (E_K)
\,, \label{eq:DM-antiproton}
\end{eqnarray}
with
\begin{eqnarray}
 R_k (E_K)
  = 
\frac{\sum_{i=1}^{\infty}  \beta_{i,k}  J_0 \big( \zeta_i \frac{r_\odot}{R_c} \big)}{ \frac{1}{2} \Big( \frac{\rho_\odot}{m_{\rm DM}} \Big)^2   \langle \sigma v \rangle_k \,   \Big(\frac{ d {\cal N}_{\bar p}}{d E_K} \Big)_k }
 \,.
\label{eq:R-EK}
\end{eqnarray}
In the zeroth order approximation ($\beta_{i,k} \to \beta_{i,k}^{(0)}$), we have 
 \begin{eqnarray}
R_k (E_K) \approx R^{(0)} (E_K) 
 = \sum_{i=1}^\infty \frac{\bar{y}_p^{i} (y_L) }{A_i \, y_1^i (y_L)} 2 D_{xx} \bigg( \frac{V_0}{2D_{xx}} \bigg)^{1/2} 
  \, J_0 \bigg( \zeta_i \frac{r_\odot}{R_c} \bigg)  \,,
 \label{eq:R}
\end{eqnarray}
which is independent of the DM mass and its annihilation channel.

\section{The result for fitting the AMS-02 data: a simplified DM model, $\text{DM~DM} \to \lowercase{\bar b}\, \lowercase{b}$,
and comparison with the GALPROP calculation
}\label{sec:simplified-dm}

Here, using the semi-analytical solution, we consider a simplified DM model, $\text{DM~DM} \to \lowercase{\bar b}\, \lowercase{b}$ to interpret the AMS-02 antiproton excess data.
We use the GALPROP WebRun \citep{Vladimirov:2010aq} to generate the proton and secondary antiproton spectra (called the background spectrum in the present work) from all cosmic-ray species, where the adopted parameter set is consistent with the model F in Ref.~\citet{Cholis:2017qlb}. Here, the radius of the cylinder $R_c=25$~kpc and half-height $L=3$~kpc are used.

Taking into account the energy-dependent uncertainty of the antiproton production in the proton-proton collision and antineutron production in the interstellar medium, where the latter can result in the antiproton production via the antineutron decay, we further parametrize the GALPROP result for the secondary antiproton flux with an additional scaling factor~\citep{Cholis:2019ejx},
\begin{eqnarray}
S_F (E_{K}^{\textrm{ISM}}) = 
a + b~\ln \bigg(\frac{E_{K}^{\textrm{ISM}}}{{\rm GeV}}\bigg) + c \bigg[\ln\bigg(\frac{E_{K}^{\textrm{ISM}}}{{\rm GeV}}\bigg)\bigg]^{2},
\label{eq:scaling}
\end{eqnarray}
which is constrained by the range of energy-dependent 3$\sigma$ uncertainties given in Ref.~\citet{diMauro:2014zea}\footnote{An analysis in Ref.~\citet{Korsmeier:2018gcy} gives  uncertainties on the source about up to $\pm 20\%$.}. Moreover, we multiply a factor $``d"$, constrained by $|d-1| \leq 0.05$,  to the antiproton flux to account for possible uncertainties of the interstellar gas density used in the GALPROP.

When protons and antiprotons propagate to enter the solar system, they will experience a process of energy loss, which is known as solar modulation, resulting in the energy shift, $E_K = E_K^{\rm ISM} - |q| \phi$, where $|q|$ is the absolute value of their charge. We use an empirical model  
for the modulation potential \citep{Cholis:2019ejx},
\begin{eqnarray}
\phi(R,t,q) &=& \phi_{0} \, \bigg( \frac{|B_{\rm tot}(t)|}{4\, {\rm nT}}\bigg) + \phi_{1} \, N'(q) 
H(-qA(t))  \bigg( \frac{|B_{\rm tot}(t)|}{4\,  {\rm nT}}\bigg) \, \bigg(\frac{1+(R/R_0)^2}{\beta (R/R_{0})^3}\bigg) 
\, \bigg( \frac{\alpha(t)}{\pi/2} \bigg)^{4}, 
\label{eq:mod-potential}
\end{eqnarray}
where $R_0\equiv 0.5$~GV, $R$ is the cosmic-ray rigidity prior to entering the Helipsphere, $|B_{\rm tot}|$ and $A$ are respectively the strength and polarity of the Heliospheric magnetic field (HMF) at the Earth, $\alpha$ is the tilt angle of the Heliospheric current sheet, $H$ is the Heaviside step function, and $N' (q) H(-qA) \in [0,1]$ depends on the HMF geometry.  In addition to the potential parameters which are allowed in the ranges, $0.32 \leq \phi_0 \leq 0.38$~GV and $0 \leq \phi_1 \leq 16$~GV in the fit, all relevant values can be found in Table II of Ref.~\citet{Cholis:2017qlb}.

We use the generalized Navarro-Frenk-White (gNFW) profile  \citep{Navarro:1995iw,Navarro:1996gj} to parametrize the DM density distribution in our Galaxy, 
\begin{eqnarray}
 \label{eq:gNFW}
 \rho(\vec{x})=\displaystyle \rho_{\odot} \left(\frac{ |\vec{x}| }{r_\odot}\right)^{-\gamma} \left(\frac{1+ |\vec{x}| /r_s}{1+r_\odot /r_s}\right)^{\gamma-3} \,,
 \end{eqnarray}
with $|\vec{x}| =(r^2+ z^2)^{1/2}$ (where $r$ is the Galactocentric radius, and $z$ is the distance from the Galactic disc), $r_s=20$~kpc and  $\rho_\odot =0.4~\text{GeV}/\text{cm}^3$ which, as a default value, is the local DM density corresponding to  $r_\odot = 8.5$~kpc.  We use $\gamma=1.2$ which is preferred by the Galactic gamma-ray excess \citep{Calore:2014xka}. 

\begin{figure}[t!]
\begin{center}
\includegraphics[width=0.45\textwidth]{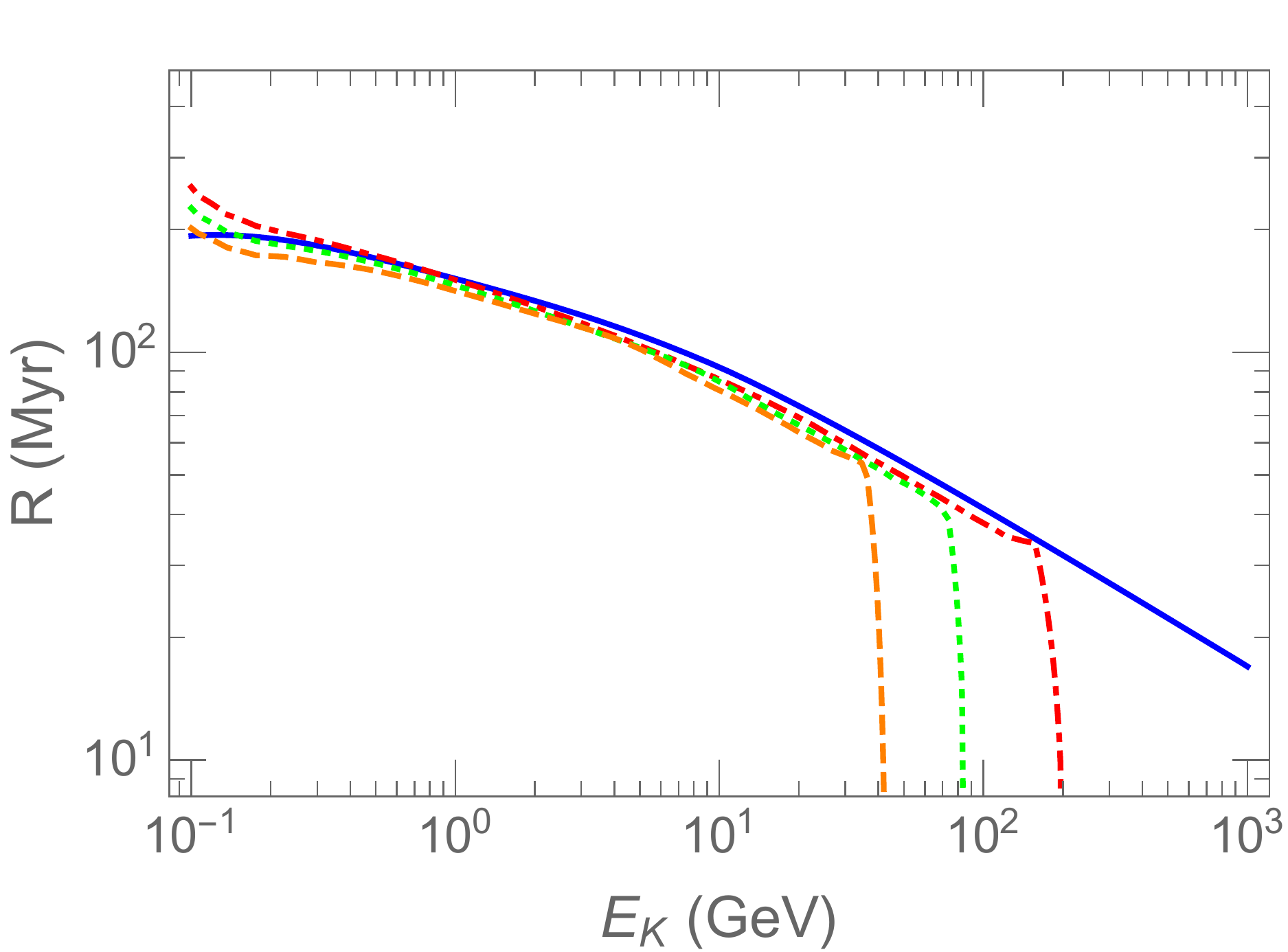}
\caption{
The propagation function $R$ of cosmic-ray antiprotons, given in Eq.~(\ref{eq:R-EK}), as a function of  the kinetic energy $E_K$, where, in comparison with $R^{(0)}$ which denotes as the solid blue line, the dotdashed (red), dotted (green), and dashed (orange) lines are the full results corresponding to $m_{\text{DM}} =200, 90$ and 50~GeV, respectively. 
\label{fig:Rratio} }
\end{center}
\end{figure}

For modeling the cosmic-ray antiprotons generated form the DM annihilation, we  need one more parameter, ``$h$" which describes the half height of the Galactic gas distribution in the two-zone  cylindrical model and is relevant to the propagation of cosmic-ray flux generated from the DM annihilation.
Using $h=0.1$~kpc,  in Fig.~\ref{fig:Rratio}, we show the propagation function $R$ of cosmic-ray antiprotons, given by Eq.~(\ref{eq:R-EK}), as a function of  the kinetic energy $E_K$. The result shows that the energy loss processes, including Coulomb, ionization and adiabatic energy losses, slightly modify $R^{(0)}$, while the tertiary mechanism, that neither creates nor annihilates antiprotons, redistributes them towards the low energy region with $E_K/m_{\text DM}\lesssim 0.003$.  In the next section, we will further discuss the dependence of our results on the value of $h$.

\begin{table*}[t]
\begin{adjustwidth}{-1.5cm}{}
\resizebox{\textwidth}{!}{
    \begin{tabular}{ccccccccccc}
         \hline\hline
      Parameter & $m_\text{DM}$  & ~$\langle \sigma v\rangle$ & ~$a$ & ~$b$ &  ~$c$& ~ $d$ & ~$\phi_0$ &  ~$\phi_1$ & $\chi_{\rm min}/{\rm dof}$ & p-value \\
            (unit)  &         (GeV)         & ~~~$(10^{-26}~\text{cm}^3/s)$ &  ---   &   ---    &   ---  &   ---   &  GV & GV & --- & ---\\
            \hline
 \multicolumn{10}{c}{ {\color{red} 4-year data set (from May 2011 to 2015) of AMS-02~\citep{Aguilar:2016kjl} with \# of data points $=57$ }} \\
            fitting without DM &  --- & ---    & ~1.147 & ~$-0.165$ & ~0.029 & ~1.034 & ~0.364 & ~0.0001  & 58.8/51 & 0.21 \\
           fitting  DM            & 87.1& 3.16 & ~1.051 & ~$-0.170$ & ~0.036& ~1.023 & ~0.320 & ~0.0000   & 36.6/49 & 0.90 \\
    \hline\hline
        \end{tabular}}
\caption{Best-fit results. Here we use $h=0.1$~kpc. \label{tab:best-fit}} 
 \end{adjustwidth}
\end{table*}

 We summarize the results for the best $\chi^2$ fit to the AMS-02 $\bar{p}/p$ data\footnote{The AMS-02 data are obtained from the Cosmic Ray DataBase (CRDB) \citep{DiFelice:2017ahk} shown at the website: ``https://tools.ssdc.asi.it/CosmicRays/".}
 in Table~\ref{tab:best-fit}, where the fit is performed in the 6-dimensional parameter space $(a, b, c, d, \phi_0, \phi_1)$ or  8-dimensional parameter space $(m_{\rm DM}, \langle \sigma v\rangle, a, b, c, d, \phi_0, \phi_1)$ for the case without or with contributions arising from the DM annihilation to the $\bar{p}$ spectrum \citep{Aguilar:2016kjl}.
 The corresponding spectra, together with the data points, are shown in Fig.~\ref{fig:fits-2016}. 
In Fig.~\ref{fig:dm-comparison}, using the semi-analytical formula given in Eqs.~(\ref{eq:flux}) and (\ref{eq:DM-antiproton}),  we further show the corresponding best-fit result for the antiproton flux, produced from ${\rm DM~DM} \to \bar{b} \, b$, at the solar location but before undergoing the effect of solar modulation, in comparison with that obtained from the GALPROP approach (see also footnote 9). Although a simplified ISM distribution is used in the semi-analytical method, and however a spatial dependent ISM distribution is adopted in the GALPROP,  the results from the two approaches are well consistent with each other for $E_K\gtrsim 1 $~GeV. The reason is due to the fact that the diffusive antiprotons with a larger kinetic energy detected at the Earth mainly come from the nearby antiproton sources.

Treating the antiprotons as a hard core gas and neglecting the Galactic wind and fragmentation, their mean free path in the diffusion process is about $3D_{xx}/v$\, \footnote{
See the Feynman lectures on physics Vol. I, Ch 43 at http://https://www.feynmanlectures.caltech.edu/I\_43.html. }. It means that
a detected antiproton with a lower kinetic energy experiences a longer propagation distance and is more sensitive to the nonlocal structure of the ISM.
It can thus be understood that the difference of the two results becomes larger in the lower kinetic energy region $E_K\sim 0.1$~GeV, as shown in Fig.~\ref{fig:dm-comparison}.
This is also supported by propagation function $R^{(0)}$ given in Fig.~\ref{fig:R-diff-h}, where a lower value of $R^{(0)}$  corresponds to 	a larger $h$ in the low kinetic energy region.

Therefore, although the GALPROP has a more realistic description about the interstellar gas distribution, where the antiprotons may be destroyed when passing through it, such the distribution can well correspond to the half-height $h$ of the Galactic disc, of order of $h\sim 0.1$~kpc, in a two-zone cylindrical model, as shown in the present study.  The formula of the semi-analytical solution for cosmic-ray antiprotons produced from the secondary sources is presented in the Appendix~\ref{app:discrete-sources}.  A consistent version for the code of the linear wind two-zone cylindrical model is thus possible. I will leave it as a future work.

\begin{figure}[t!]
\begin{center}
\includegraphics[width=0.38\textwidth]{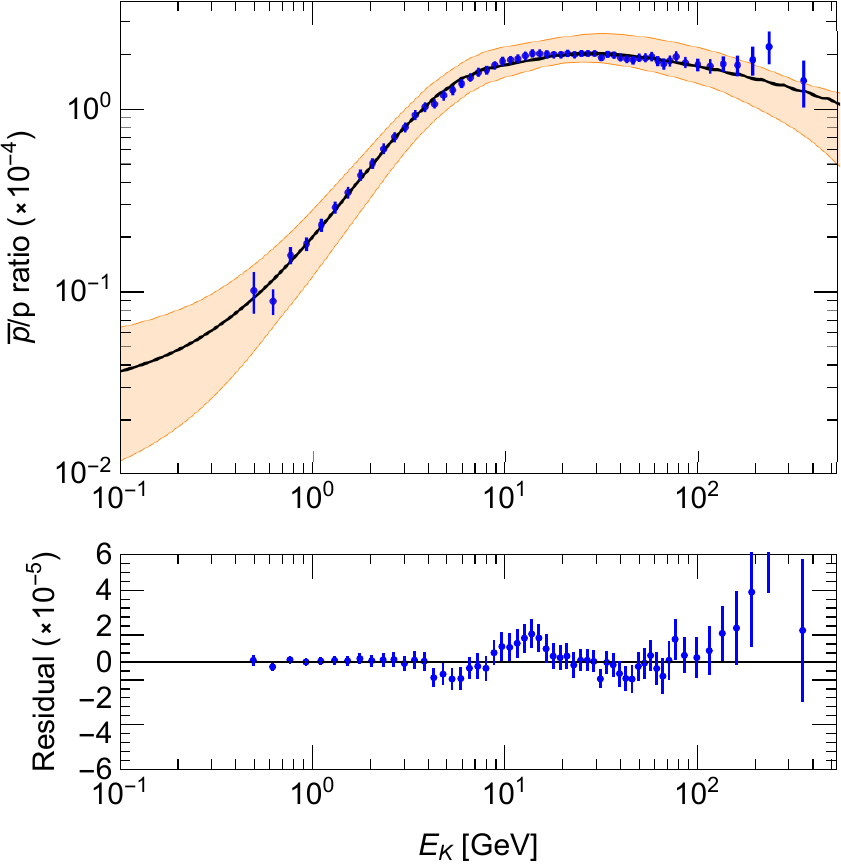}\hskip0.6cm
\includegraphics[width=0.38\textwidth]{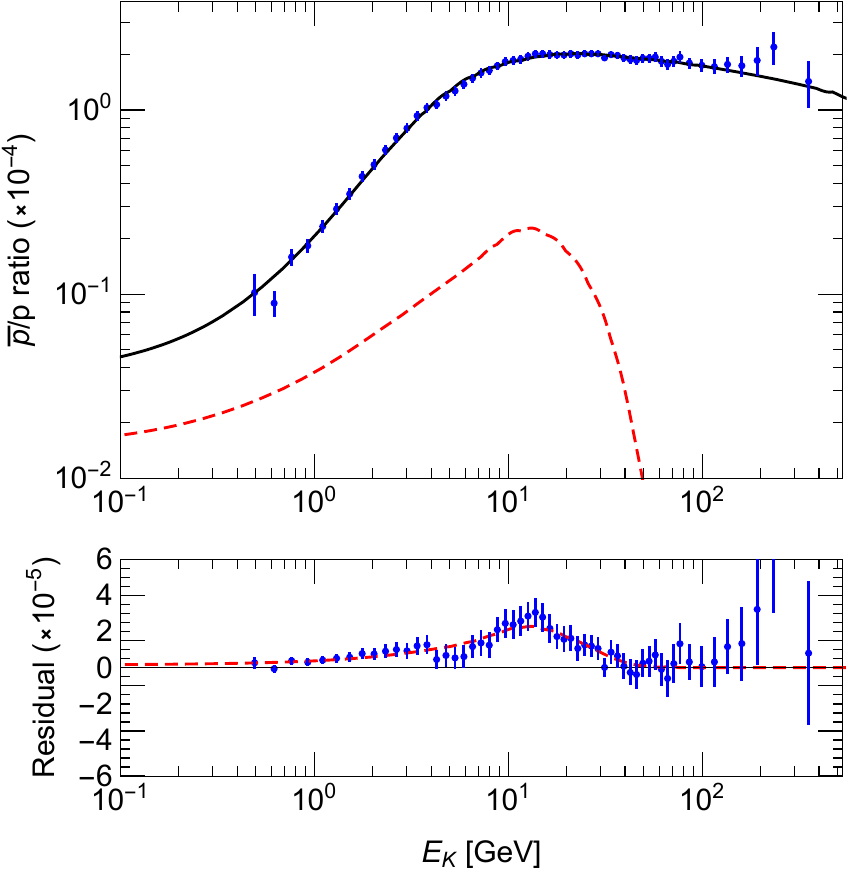}
\caption{  Left panel: The best fit ${\bar p}/p$ without contribution from DM annihilation (black solid line) vs. 4-year data set collected by AMS-02 from May 2011 to May 2015 \citep{Aguilar:2016kjl},  where the range in orange color is allowed due to variations of parameters within the possible uncertainties, for which $a, b$ and $c$ are constrained by energy-dependent 3$\sigma$ uncertainties given in Ref.~\citet{diMauro:2014zea}, $|d-1| \leq 0.05$,  $0.32 \leq \phi_0 \leq 0.38$~GV and $0 \leq \phi_1 \leq 16$~GV. Right panel: The best-fit result (black solid line)  but including the contribution from $\text{DM~DM} \to {\bar bb}$, corresponding to $m_{\rm DM} =$ 
87.1~GeV and $\langle \sigma v\rangle=$
3.16 $\times 10^{-26}~\text{cm}^3/s$.  The contribution of the DM annihilation is depicted by the dashed (red) line. 
 The corresponding residuals with the observed data minus the astrophysical background are shown in the lower panels. \label{fig:fits-2016} }
\end{center}
\end{figure}

\begin{figure}[t!]
\begin{center}
\includegraphics[width=0.42\textwidth]{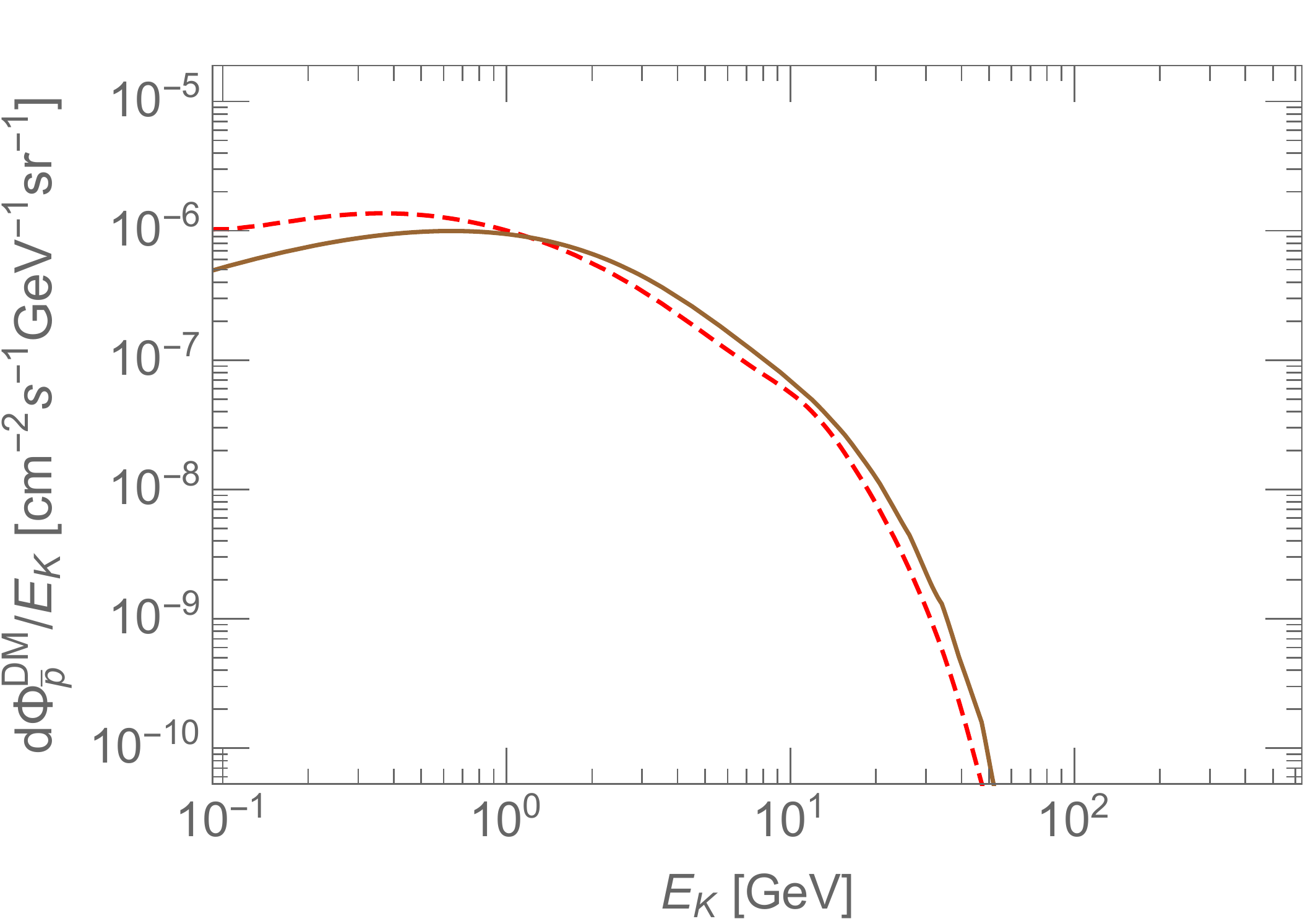}
\caption{ Comparison between the GALPROP (solid brown curve) and semi-analytical (dashed red curve) results for the antiproton flux, generated from ${\rm DM~DM} \to \bar{b} \, b$ with $m_{\rm DM}$=87.1 GeV, at the solar location but before undergoing the effect of solar modulation. In the semi-analytical result, $h=1$~kpc is used.  
\label{fig:dm-comparison} }
\end{center}
\end{figure}

The upper limit (UL) on $\langle \sigma v \rangle $ and favored confidence region on the  $ (m_{\rm DM}, \langle \sigma v \rangle) $ plane are shown in Appendix~\ref{app:contours}.
Our result shows that the parameter range of $\text{DM~DM} \to {\bar b \, b}$ indicated by AMS-02 data prefers a larger DM mass compared with those obtained from the analysis of the Galactic center gamma-ray excess \citep{Calore:2014xka,Yang:2017zor}. It should be noted that the parameter range of the DM model in the analysis of  AMS-02 data depends on the parameter set used in the diffusion-convection equation. On the other hand, one should also note that in, {\it e.g.}, Ref. \citet{Heisig:2020nse}, the authors have argued that systematic uncertainties can significantly reduce the significance of the antiproton excess. 
However, a detailed study for these parts is beyond the scope of this paper.

\section{ {\lowercase{\it h}}-dependence of the contribution arising from DM annihilation}\label{sec:h-dep}

\begin{figure}[t!]
\begin{center}
\includegraphics[width=0.38\textwidth]{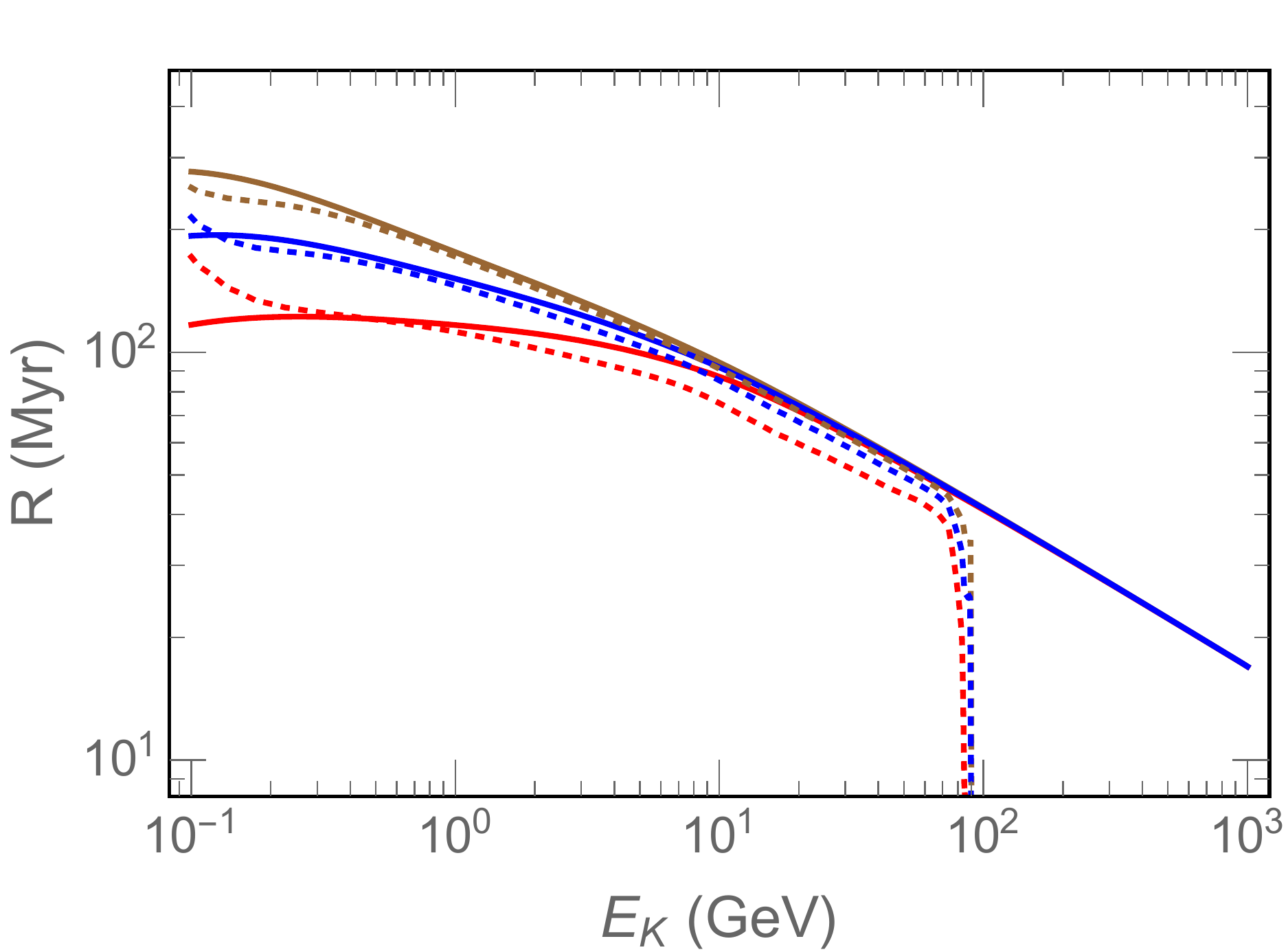}
\caption{
The propagation function $R$ of the cosmic antiprotons, produced from DM annihilation, as a function of  the kinetic energy $E_K$, where at $E_K=1$~GeV from up to down  the solid lines (with color in brown, blue and red) stand for $R^{(0)}$ corresponding to the use of $h= 0.05, 0.1$ and 0.2~kpc, respectively, while the dotted lines denote the full results of $R$ with $m_{\rm DM} =90$~GeV. \label{fig:R-diff-h}
}
\end{center}
\end{figure}

Comparing the present work with the cylindrically symmetric GALPROP model, where a radial dependence of the interstellar hydrogen distribution is considered, to find a semi-analytical formula for the cosmic antiproton density produced from the DM annihilation we have modeled that the interstellar matter uniformly distributes in the narrow disc with a half height $h$ in the Galaxy.

In Sec.~\ref{sec:simplified-dm}, we have used $h=0.1$~kpc
in the analysis.  Here we discuss the $h$ dependence of the propagation function of cosmic antiprotons produced by DM annihilation. Using three different values, $h=0.05, 0.1$, and 0.2~kpc, the propagation function of cosmic-ray antiprotons is displayed in Fig.~\ref{fig:R-diff-h}, where the three $R^{(0)}$ curves are distinct for $E_K \lesssim10$~GeV, and becomes notably different in value for $E_K \lesssim 2$~GeV as the process for antiprotons annihilating on the interstellar protons dominates  at the low energy. 
If including the energy loss and diffusive reaccelaration terms in the calculation, $R$ is modified and becomes smaller for a larger $h$ (compared with $R^{(0)}$). Meanwhile, for the case of cosmic antiprotons, one needs to further include the tertiary contribution, which is treated as a source term, for a full consideration.  As shown in Fig.~\ref{fig:R-diff-h}, the tertiary antiprotons
(see also footnote 2)
are significantly enhanced at low energy for a larger $h$, so that the propagation function is also manifestly redistributed towards the low energy region with $E_K/m_{\text DM}\lesssim 0.003$.

However, as seen in the right panel of Fig.~\ref{fig:fits-2016},  the best fit shows that the contribution from the DM annihilation to the antiproton spectrum is only about 10\% of that from the secondary production at the low energy,  $E_K \lesssim 2$~GeV. As a result, for $m_{\rm DM}\gtrsim 60$~GeV, the fitting value of $\langle \sigma v \rangle $  depends on $h$ but is not so sensitive to its value change. As a comparison, in Fig.~\ref{fig:ams-contours-LH} of Appendix~\ref{app:contours}, we show the contours on the $ (m_{\rm DM}, \langle \sigma v \rangle) $ plane using $h=0.05$ and 0.2 kpc.

\section{Remarks on the advantage of the semi-analytical approach} \label{sec:comp-adv}

The advantages for using the semi-analytical approach to the study of antiprotons produced by DM annihilation are given as follows. 

Compared with the GALPROP, which  based on the finite difference scheme needs huge memory for higher resolutions and takes more computation time, the semi-analytical approach can be much faster than the numerical models in computation and easily avoid the numerical instability. 

To compare the computing time of two approaches, we perform the numerical calculations on a Linux OS installed via Parallels virtual machine on a Mac desktop with 4 GHz Intel Core i7 and 8GB 1600MHz LPDDR3. 
Using the GALPROP v54 code\footnote{
We use the version v54.r2766 available at: https://gitlab.mpcdf.mpg.de/aws/galprop. A DM extension code by Andrey Egorov is available at https://github.com/a-e-egorov/GALPROP\_DM. Compared with the original code by Egorov, here, (i) the tertiary contribution is further included and (ii),  instead of introducing a truncation radius for the gNFW profile to treat the unphysical divergence, the profile in the grid cells nearby the Galactic  center is taken to be its value at $r=dr/2$ and $z=dz/2$, so that the numerical output is faster and more stable.}, 
where the grid step $(dz ,dr)=(0.1~{\rm kpc}, 0.25~{\rm kpc}$) was used, and the halo profile was then averaged over 1/10 size of the grid size,
 we spent about 3 mins for each DM mass input to get an output for the antiproton flux produced from the DM annihilation with 54 points in the energy spectrum.
 It thus needs about 150 mins to have $50\times 54$ points on the $(m_{\rm DM}$, $E_K)$ plane, where $E_K <m_{\rm DM}$ and $m_{\rm DM} \in$ (5.2-1000)~GeV.  
Up to a factor, this antiproton flux result is equivalent to the propagation function $R$ in the semi-analytical approach.

By running the C$^{++}$ program, 
this corresponding value of $R$ with the same number of points 
$(50\times54)$ on the $(m_{\rm DM}$, $E_K)$ plane
can be obtained in 
less than 120
seconds for having a stable solution, where the first 
46
terms in the Bessel expansion series (defined in Eq.~(\ref{eq:multi-expansion})) have been computed. Since the semi-analytical approach has a faster output, it can thus be positive for building a more realistic DM model for which the DM resides in a hidden sector with more particles and couplings involved. 

This approach is also useful to understand the underlying physics and make a quick estimate on the physical quantity. For instance, if the propagation function is scaled by a factor, the DM annihilation cross section is inversely proportional to it (see Eq.~(\ref{eq:DM-antiproton})).  Therefore, as shown in Fig.~\ref{fig:R-diff}, because in the  most range of kinetic energy, especially at $E_K=10~$GeV, the propagation functions $R^{(0)}$ at $\gamma=1.0$  is decreased by a factor of 0.71 compared with the value at $\gamma=1.2$,  the annihilation cross section at $\gamma=1.0$ is thus expectedly increased by a factor of $1/0.71 =1.4$ compared with that at $\gamma=1.2$. This estimate is quite close to the true factor ($=1.3$) in the best fit.

\begin{figure}[t!]
\begin{center}
\includegraphics[width=0.38\textwidth]{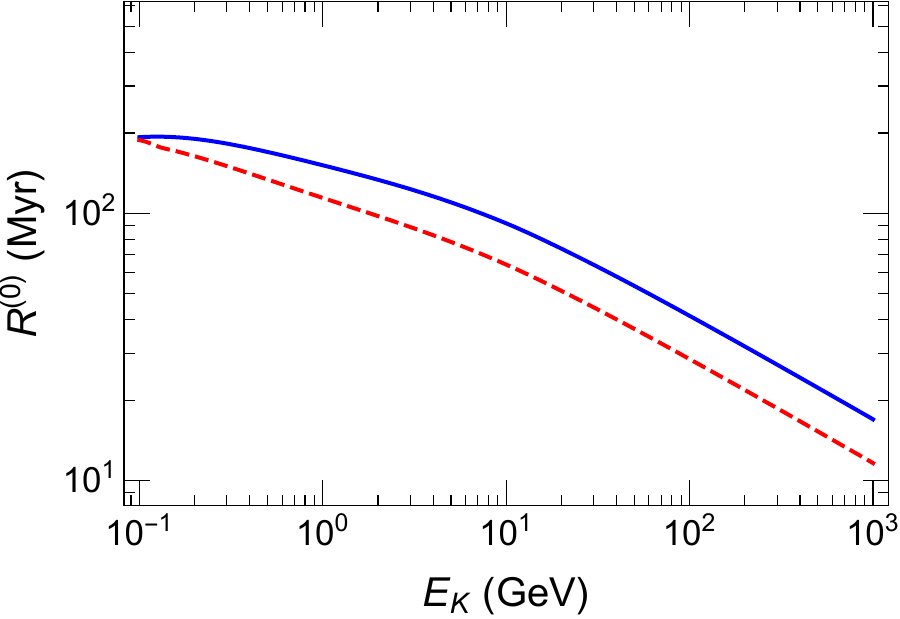}
\caption{
 The propagation function $R^{(0)}$ of the cosmic antiprotons produced from DM annihilation, as a function of  the kinetic energy $E_K$, where the blue solid and red dashed lines correspond to the use of $\gamma= 1.2$ and 1.0, respectively. The values of rest input parameters are the same as that given in Sec.~\ref{sec:simplified-dm}.}
\label{fig:R-diff}
\end{center}
\end{figure}

\section{Summary}\label{sec:summary}

In the framework of a cylindrical diffusion model taking into account a linear convective wind, the cosmic-ray propagation can be  solved using the public GALPROP code which  aims at a fully numerical solution. Alternatively, in this paper, we have offered a semi-analytical approach to obtain the spectrum of cosmic-ray nuclei produced from the primary sources throughout the dark matter halo, instead of a purely numerical way.

Considering the two-zone cylindrical model, where a linear convective wind is presented outwards from the Galactic plane, we have obtained a semi-analytic solution for the spectrum of the cosmic  antiproton density, produced from primary sources which are generated by DM annihilation/decay throughout the diffusive halo. 
The result is given in Eq.~(\ref{eq:DM-antiproton}). While the secondary antiprotons are generated using, {\it e.g.},  the GALPROP WebRun and well constrained by other observations, this semi-analytical formula will be helpful for particle physicists, who are not familiar with the details of this code,  to build a suited DM model and then to perform a statistical analysis when confronting with data. 
 Numerically, the semi-analytical approach can be much faster than the fully numerical models in computation and easily avoid the numerical instability.
Thus, the obtained semi-analytical solution for cosmic-ray antiprotons not only offers an intuitive insight on the dark matter search,  but also enables to efficiently deal with a realistic DM model with more fundamental particles involved in the interactions.

 Using the solution obtained for the antiproton propagation, we have studied the constraint on the  DM signal, through the channel $\text{DM~DM} \to {\bar b}b$, from the measurement of the AMS-02 antiproton-to-proton ratio. 
We find an indication of the DM signal with the best-fit result:  $m_{\rm DM} =$
87.1~GeV and  $\langle \sigma v\rangle=$
3.16 $\times 10^{-26}~\text{cm}^3/s$ corresponding to the $p$-value$=$
 0.90. We show that the semi-analytical result of the antiproton flux generated from ${\rm DM~DM} \to \bar{b} b$ at the solar location is well consistent with that obtained by the GALPROP approach in the kinetic energy region $E_K\gtrsim 1 $~GeV. }

The value of $h$, modelig the half height of the Galactic gas distribution in the two-zone cylindrical model, is related to the propagation of cosmic antiprotons produced by DM annihilation. 
In Sec.~\ref{sec:h-dep}, we have shown that  the fit result is insensitive to using $h\lesssim 0.1$~kpc,
while for a value change from $h=0.1$~kpc to $0.2$~kpc, the best-fit point together with the contour is approximately moved upwards on the $ (m_{\rm DM}, \langle \sigma v \rangle) $ plane with the value of  $ \langle \sigma v \rangle$ increased by a factor of 1.25 
(see also Fig.~\ref{fig:ams-contours-LH}).

\acknowledgments \vspace*{-1ex}
The author is grateful to Andrew W. Strong for useful communication.
 This work was supported in part by the Ministry of Science and Technology, Taiwan, under Grant No. 109-2112-M-033-004.

\appendix

\section{Solution of the diffusion equation for the propagation of cosmic-ray nuclei produced from background sources lying in the Galactic plane}\label{app:discrete-sources}

For the background which is related to the sources located within the half height $h$ of the Galactic disc, we represent the resulting number density $N^{(b)}$  and background source $q^{(b)}$, which satisfy the boundary conditions, $N^{(b)} (R_c,z,E_K)=0$ and $q^{(j)} (R_c,z,E_K)=0$, in terms of a complete set of orthogonal Bessel functions:
\begin{eqnarray}
N^{(b)} (r, z, E_K) &=& \sum_{i=1}^{\infty} N_i^{(b)}  (z, E_K) J_0 (\zeta_i \rho) \,,  \\
q^{(b)}(r, z, E_K) &=& \sum_{i=1}^{\infty} 2 h \delta(z) Q^{(b)}_i (E_K) J_0 (\zeta_i \rho) \,.
\end{eqnarray}
The density spectrum of the cosmic rays at the solar system is then given by
\begin{eqnarray}
N^{\rm (b)} (r_\odot, 0, E_K) 
= \sum_{i=1}^{\infty} N_i^{\rm (b)}  (0, E_K) J_0 \Big( \zeta_i \frac{r_\odot}{R_c} \Big)\,.
\label{eq:multi-expansion-b}
\end{eqnarray}
Following the procedure as done in Sec.~\ref{sec:linear-solution}, and neglecting the effects about energy losses and diffusive reacceleration, and tertiary contribution, we obtain the approximation\footnote{This approximation basically is consistent with that given in (C2) of Ref.~\citep{Taillet:2003yy}, except the factor ``$a_i$". I have checked that this factor should read
$ a_i\equiv (2K/ V_0) (\zeta_i^2/ R^2) +1$, instead of   $ a_i\equiv (2K/ V_0) (\zeta_i^2/ R^2) +2$ given in Ref.~\citep{Taillet:2003yy}.
 },
\begin{eqnarray}
N_i^{\rm (b)}  (0, E_K) \simeq 
\frac{ 2h\, Q^{(b)}(E_K)}{ A_i }  \,, 
\end{eqnarray}
where $A_i^{-1}$  have been given in Eq.~(\ref{eq:Ainv}). If considering all effects, the result is given by
\begin{eqnarray}
& &N_i^{\rm (b)}  (0, E_K)  \left[ 2 h \Gamma_{\rm ann} +2 D_{xx} \bigg( \frac{V_0}{2D_{xx}} \bigg)^{1/2}  \frac{y_2^i(y_L)}{y_1^i(y_L)} \right] 
    \nonumber\\
& & \hskip1cm 
 =  2 h Q^{(b)}  
     -2h \frac{\partial}{\partial E_K}
          \left( b_{tot} N_i^{\rm (b)}  (0, E_K) - \beta^2 D_{pp} \frac{\partial N_i^{\rm (b)}  (0, E_K)  }{\partial E_K} \right) 
   + 2h q_{i}^{\rm ter} (r_\odot, 0, E_K) \,,
  \label{eq:betai-b}
\end{eqnarray}
where the tertiary term is for the case of cosmic antiprotons.
The numerical value of $N_i^{\rm (b)}  (0, E_K)$ can be computed from Eq.~(\ref{eq:betai-b}), using the iterative procedure.

\section{ Favored and disfavored regions on the $(\lowercase{m}_{\rm DM}, \langle \sigma \lowercase{v} \rangle) $ plane}\label{app:contours}

In Fig.~\ref{fig:ams-2016-contours}, we show the contour plot on the  $ (m_{\rm DM}, \langle \sigma v \rangle) $ plane, where the upper limit on $\langle \sigma v \rangle $ is determined from  the test statistic,
\begin{eqnarray}
q =-2 
\ln \left[ \frac{ \mathcal{L} ( \langle \sigma v\rangle,  \hat{\hat{\theta}} ; m_{\rm DM} ) }{ \mathcal{L} (0 , \hat{\bf \theta} ; m_{\rm DM} ) }\right]  \,,
\label{eq:L-UL}
\end{eqnarray}
with $\theta \equiv (a, b, c, d, \phi_0, \phi_1) $ being a generic set of the nuisance parameters. Here, $\hat{\hat{\theta}}$ and  $\hat{\theta}$, called the maximum likelihood estimators (MLEs),  maximize the likelihood ${\cal L}$ for a given value of the DM annihilation cross section $\langle \sigma v \rangle $ and for the null DM measurement, respectively.
For one-side 95\% UL, we take $q=2.71$. 
To determine the confidence region for the model of DM annihilating to ${\bar b} b$, we use the log-likelihood test statistic,
\begin{eqnarray}
t =-2 
\ln \left[ \frac{ \mathcal{L} ( \langle \sigma v\rangle, m_{\rm DM},  \hat{\hat{\theta}} ) }{ \mathcal{L} ( \widehat{\langle \sigma v\rangle}, \widehat{m}_{\rm DM} , \hat{\bf \theta} ) }\right]  \,,
\end{eqnarray}
where $\hat{\hat{\theta}}$ are the MLEs for given values of $m_{\rm DM}$ and $\langle \sigma v\rangle$, and result in systematic uncertainties, while in the denominator, the parameters with a ``hat" represent MLEs corresponding to the $\chi_{\rm min}^2$ fit. In Fig.~\ref{fig:ams-2016-contours}, the  $1\sigma, 2\sigma$ and $3\sigma$  confidence regions correspond to $t-\chi_{\rm min}= 2.30, 6.18$, and 11.83, respectively. For the likelihood-based statistical tests, we refer the reader to Refs.~\citet{Cowan:2010js,Cowan:2018lhq} and references therein.

\begin{figure}[t!]
\begin{center}
\includegraphics[width=0.38\textwidth]{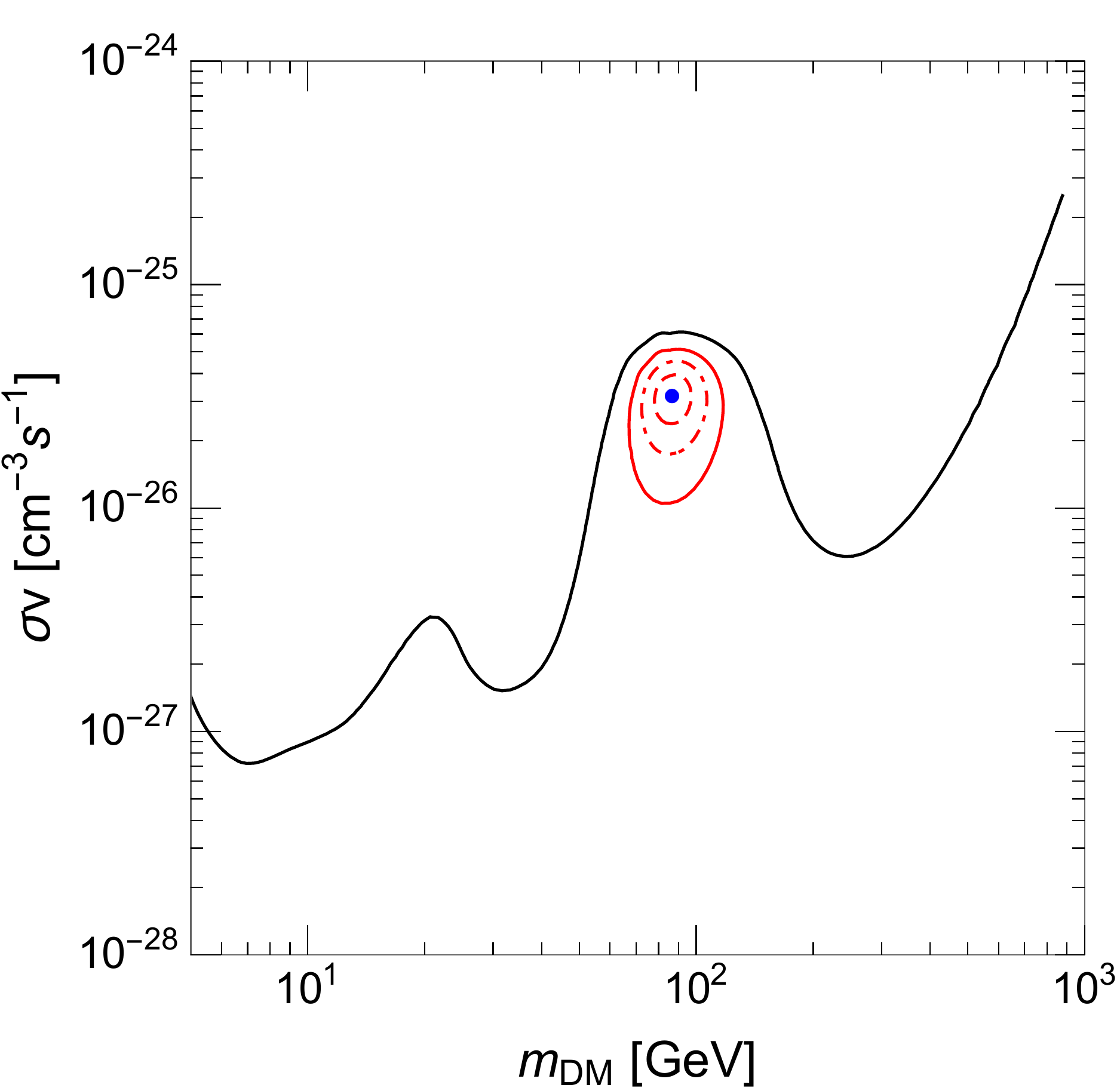}
\caption{ Favored and disfavored regions on the  $ (m_{\rm DM}, \langle \sigma v \rangle) $ plane for $\text{DM~DM} \to {\bar b} b$ through fitting to the 4-year data set collected by AMS-02 from May 2011 to May 2015 \citep{Aguilar:2016kjl}. The best fit is denoted by the blue dot, and the  confidence regions within $1\sigma, 2\sigma$ and $3\sigma$ are bounded by dashed, dotdashed, and solid curves. The 95\% UL, defined through Eq.~(\ref{eq:L-UL}) with $q=2.71$, is shown by the black line.
}
\label{fig:ams-2016-contours}
\end{center}
\end{figure}

As a comparison, in Fig.~\ref{fig:ams-contours-LH}, we show the contours on the $ (m_{\rm DM}, \langle \sigma v \rangle) $ plane using $h=0.05$ and 0.2 kpc, where the best fit points are $ (m_{\rm DM}, \langle \sigma v \rangle) =(87.4~\text{GeV}, 2.96\times 10^{-26}~\text{cm}^3/s)$ and $(86.8~\text{GeV}, 3.62\times 10^{-26}~\text{cm}^3/s)$, respectively. Basically, the best-fit value of the DM mass is insensitive to a value of $h \lesssim 0.1$~kpc, while changing $h$ from 0.1~kpc to 0.2~kpc, the best-fit contour lines are approximately moved upwards on the $ (m_{\rm DM}, \langle \sigma v \rangle) $ plane with 
the corresponding value of  $ \langle \sigma v \rangle$ increased by a factor of 1.25.

\begin{figure}[t!]
\begin{center}
\includegraphics[width=0.38\textwidth]{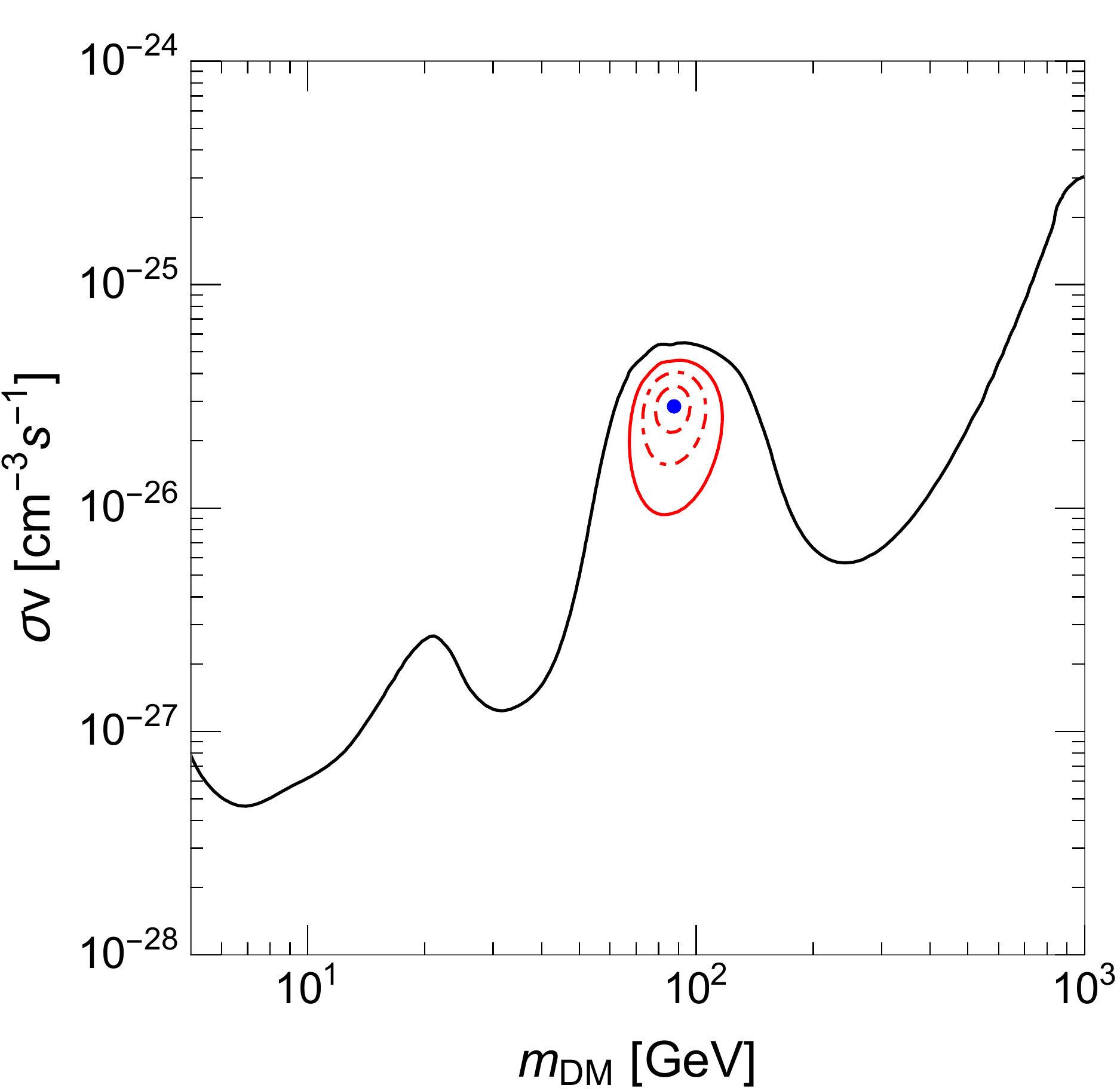}\hskip0.6cm
\includegraphics[width=0.38\textwidth]{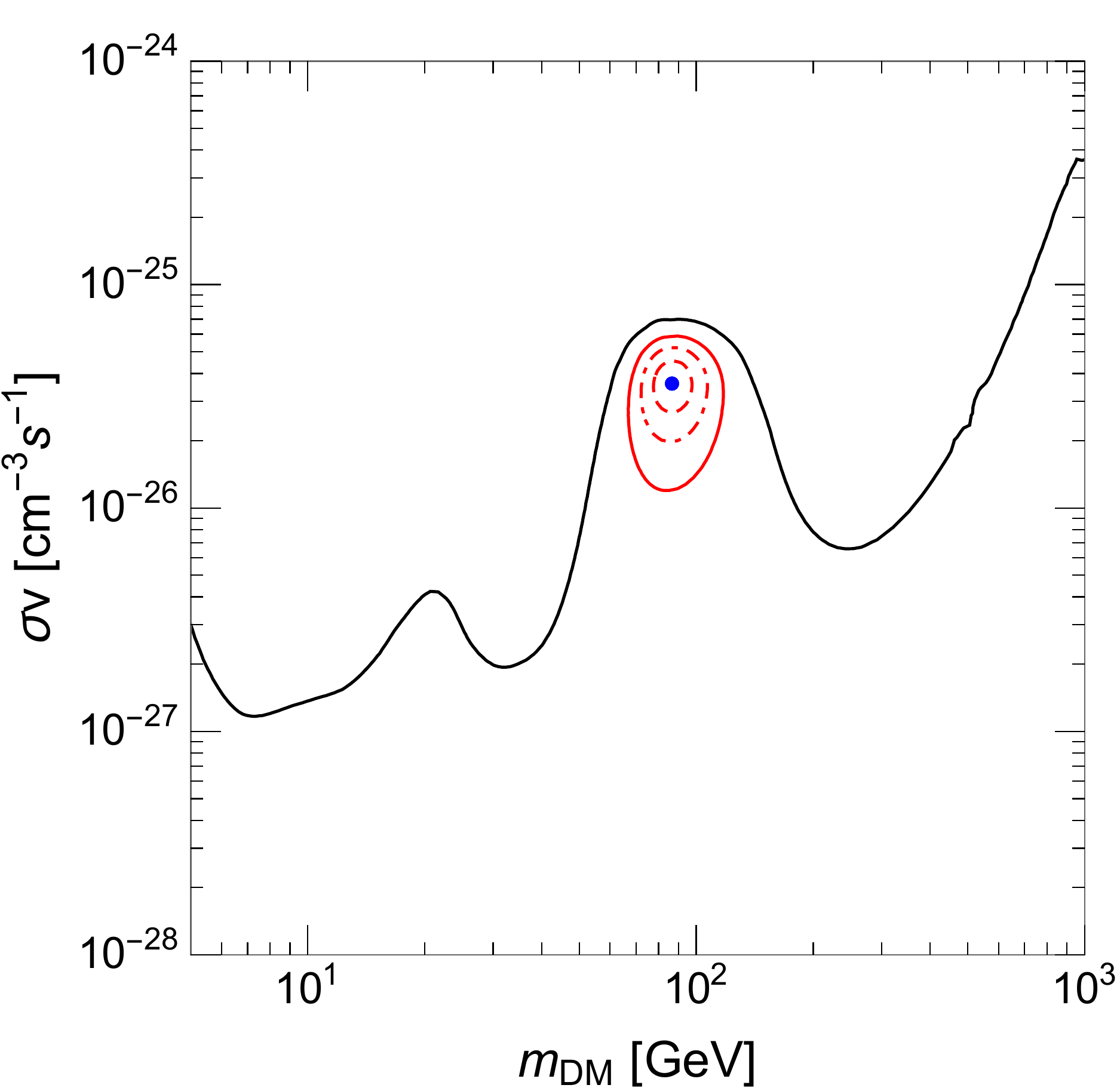}\caption{ As in Fig.~\ref{fig:ams-2016-contours} but $h=0.05$~kpc and $h=0.2$~kpc for the left and right panels, respectively. In the left panel, the best fit point is $ (m_{\rm DM}, \langle \sigma v \rangle) =(87.4~\text{GeV}, 2.96\times 10^{-26}~\text{cm}^3/s)$ corresponding to $\chi_{\rm min}= 37.0$, while in the left panel, $ (m_{\rm DM}, \langle \sigma v \rangle) =(86.8~\text{GeV}, 3.62\times 10^{-26}~\text{cm}^3/s)$ corresponding to $\chi_{\rm min}= 36.5$.
}
\label{fig:ams-contours-LH}
\end{center}
\end{figure}

\newpage



\end{document}